\documentclass[aps,preprint]{revtex4}

\def\la{\langle}
\def\ra{\rangle}

\def\lb{\lbrack}
\def\rb{\rbrack}

 \def\be{\begin{eqnarray}}
\def\ee{\end{eqnarray}}

\def\bpsi{\bar{\psi}}

\def\bpsi{\bar{\psi}}

\usepackage{graphicx}

\begin{document}

\title{Chiral Limit of Strongly Coupled Lattice Gauge Theories}
\author{David H. Adams$^{a,b}$ and Shailesh Chandrasekharan$^a$}
\affiliation{
$^a$ Department of Physics, Box 90305, Duke University,
Durham, North Carolina 27708, USA \\
and \\
$^b$ Instituut-Lorentz for Theoretical Physics, Universiteit Leiden,
Nield Bohrweg 2, 2300 RA Leiden, The Netherlands.}

\preprint{DUKE-TH-03-237, INLO-PUB-02/03}

\begin{abstract}
We construct a new and efficient cluster algorithm for updating strongly 
coupled $U(N)$ lattice gauge theories with staggered fermions in the chiral 
limit. The algorithm uses the constrained monomer-dimer representation of the 
theory and should also be of interest to researchers working on other 
models with similar constraints. Using the new algorithm we address questions 
related to the chiral limit of strongly coupled $U(N)$ gauge theories 
beyond the mean field approximation. We show that the infinite volume chiral 
condensate is non-zero in three and four dimensions. However, on a square 
lattice of size $L$ we find 
$\sum_x \ \langle \bar\psi\psi(x)\ \bar\psi\psi(0) \rangle 
\sim L^{2-\eta}$ for large $L$ where $\eta = 0.420(3)/N + 0.078(4)/N^2$.
These results differ from an earlier conclusion obtained using a different 
algorithm. Here we argue that the earlier calculations were misleading 
due to uncontrolled autocorrelation times encountered by the previous 
algorithm.
\end{abstract} 

\maketitle

\section{INTRODUCTION}

One of the challenges in the study of strong interactions is to 
compute physical quantities with in the framework of QCD with controlled
errors. Although lattice QCD can in principle accomplish this task,
most known algorithms encounter problems related to critical slowing 
down as the quark masses become small \cite{Ber02}. It is impossible
to compute quantities reliably for realistic ``up'' and ``down''
quark masses with current algorithms. At present, one usually computes 
quantities at an unphysically large quark mass and uses chiral 
extrapolations to obtain the final answer \cite{Ber102}. 
For sufficiently light quarks this is a reliable technique. However,
for the quark masses that are currently used, such extrapolations are 
questionable. Furthermore, most calculations with light quarks are obtained 
in the quenched approximation where the internal dynamics of fermions 
are ignored. There also exist some interesting quantities which 
cannot be computed when the quarks are heavy. For example, in the real 
world the rho meson can decay into two pions. Unfortunately, for most 
values of the quark masses that are currently used, this decay is 
forbidden. If lattice simulations were possible with sufficiently light
quarks, the resonant nature of the rho meson can be studied 
with lattice techniques \cite{Lus91}.

There are two main problems in constructing efficient algorithms 
for lattice QCD. The first is that the fundamental variables, namely
the quark and gluon fields, are unphysical since they are gauge 
dependent. Secondly, the quarks are fermions and due to the Pauli
principle introduce sign problems which are typically very difficult 
to solve. Fortunately, with a good lattice fermion formulation and 
at zero baryon chemical potential, the second problem can be avoided. 
The quarks can be integrated out and their effects can be encoded in 
terms of a determinant which is a positive non-local function 
of the gauge fields. However, one is still left with the problem 
of updating the unphysical gauge fields. Previous experience suggests
that in order to find an efficient algorithm it is often useful
to understand and isolate the physical modes of the theory. Unfortunately,
for gauge theories, especially with the non-local fermion determinant 
this problem looks formidable.

Interestingly, there is a limit of lattice QCD with staggered fermions
where most of the above mentioned complications can be eliminated. This is 
the so called strong coupling limit \cite{Dro83}. Although this limit
has the worst lattice artifacts and could describe the wrong phase,
the resulting theory is still an interesting toy model for QCD. It contains 
the physics of confinement and chiral symmetry breaking. In the chiral limit 
one finds massless Goldstone bosons in addition to other massive mesonic 
and baryonic excitations. The strong coupling theory can be solved 
analytically only in the large $N$ \cite{Kaw81} and large $d$ \cite{Klu83} 
limits. One needs a numerical approach to study it without these 
approximations.

The luxury of the strong coupling limit is that it is possible to 
integrate out the gauge fields analytically and write the problem entirely in
terms of gauge invariant objects. It was pointed out in \cite{Wol84} that 
with $U(N)$ gauge fields the strong coupling model with staggered fermions
is equivalent to a system of monomers and dimers with positive  
Boltzmann weights. Later a monomer-dimer-polymer representation was 
discovered for $SU(3)$ gauge fields \cite{Kar89}. Is it possible to use
these simplified representations that arise at strong couplings to 
construct an efficient algorithm in the chiral limit? In \cite{Wol84,Kar89}
simple local algorithms were proposed. However, as far as we know, all these 
algorithms break down in the chiral limit. In fact there is evidence that 
the proposed algorithms may also have other problems 
away from the chiral limit \cite{Alo00}. Recently, a cluster based algorithm 
was proposed to update the monomer-dimer system which works very well in 
the chiral limit on small lattices \cite{Cha01,Cha02}. However, even this 
algorithm becomes inefficient for large lattice sizes due to long auto 
correlation times.

In this paper we propose a new and efficient algorithm to study the chiral 
limit of strongly coupled $U(N)$ gauge theories with staggered fermions in 
any dimension. It should be possible to extend it to $SU(N)$ gauge theories 
with minor modifications. We then use this algorithm to study questions 
related to chiral symmetry in two, three and four dimensions. Our study
shows that this algorithm has the potential to address many unresolved 
questions about the chiral limit of gauge theories at least in the strong 
coupling limit. Our paper is organized as follows. In section 2 
we review the monomer-dimer representation of strong coupling $U(N)$ gauge 
theories with staggered fermions and discuss the consequences of 
chiral symmetry in this language. We then show how a finite size scaling
formula for the chiral susceptibility can be used to compute the 
chiral condensate from finite volume lattice calculations in the 
chiral limit. In section 3 we construct a new type of cluster algorithm,
which we call the ``directed path'' cluster algorithm, to 
update the monomer-dimer model. We show that it obeys detailed balance
and is ergodic. Section 4 contains explicit expressions that can be
used to compute observables like the condensate and the 
susceptibility with the new algorithm. In section 5 we test the
algorithm by comparing exact results on small lattices with the 
results obtained using the algorithm. We also present an exact large L 
result on the 2 x L lattice and compare with the result from the 
algorithm. In section 6 we discuss the 
performance of the new algorithm and compare it with an earlier 
algorithm proposed in \cite{Cha01,Cha02}. We argue why 
the results obtained earlier were incorrect at large volumes. In 
this context we also briefly study the autocorrelations of the new 
algorithm. In section 7, using our new algorithm we study the issue 
of chiral symmetry breaking in two, three and four dimensions for 
different values of $N$. Section 8 contains our conclusions.

\section{MONOMER-DIMER REPRESENTATION}

Let us review the monomer-dimer representation of the strongly
coupled $U(N)$ lattice gauge theory with staggered fermions. The 
Euclidean space action of the model we consider is given by
\begin{equation}
\label{fact}
S \ = \ - \sum_{x,\mu} \eta_\mu(x)\Big[
\bar\psi(x)U_\mu(x)\psi(x+\hat{\mu})
- \bar\psi(x+\hat{\mu})U^\dagger_\mu(x)\psi(x)\Big]
- m \sum_x \bar\psi(x)\psi(x)
\end{equation}
where $x\equiv(x_1,x_2,...,x_d)$ labels the sites on a d-dimensional
periodic, hyper-cubic lattice and $\mu=1,2,..,d$ labels the various 
directions. For concreteness we assume $x_\mu \in  0,1,2,...,L_\mu-1$ such 
that $L_\mu$ is the length of the hyper-cubical box in the $\mu$
direction. We will use $V= L_1 L_2\cdot \cdot L_d$ to denote the volume
of the lattice. The site next to $x$ in the 
positive $\mu$ direction is labeled $x+\hat{\mu}$. The link variables 
connecting the sites $x$ and $x+\hat{\mu}$, represented 
by $U_\mu(x)$, are $N\times N$ unitary matrices. $\psi(x)$ is an $N$ 
component column vector and $\bar\psi(x)$ is an $N$ 
component row vector. Both these vectors are made with Grassmann 
variables and represent the staggered fermion fields at the 
site $x$. We will assume that the gauge links satisfy periodic boundary 
conditions while the fermion fields satisfy either periodic or 
anti-periodic boundary conditions. The factors $\eta_\mu(x)$ are the
well known staggered fermion phase factors. However, we will choose them 
to be $\eta_1(x) = t$ and $\eta_\mu(x) = \exp[i\pi(x_1+..+x_{\mu-1})], 
\mu = 2,3,...d$, where $t$ is a real-valued parameter incorporating
the effects of temperature. By working on asymmetric lattices with 
$L_1 << L_\mu, \mu=2,3..d$  and allowing $t$ to vary continuously,
we can study finite temperature phase transitions \cite{Boy92}. When 
$t=1$, the $\eta_\mu(t)$ turn into the usual phase factors. The 
parameter $m$ controls the fermion mass. Our 
definition of the action, (eq.(\ref{fact})), differs from the conventional 
definition since we have dropped a factor of half in front of the 
kinetic (hopping) term. This only changes the normalization of the fermion 
fields and the definition of the fermion mass $m$ up to a factor of $2$,
but helps in avoiding extra powers of two in many expressions
we write below. 

\begin{figure}[htb]
\begin{center}
\includegraphics[width=25pc]{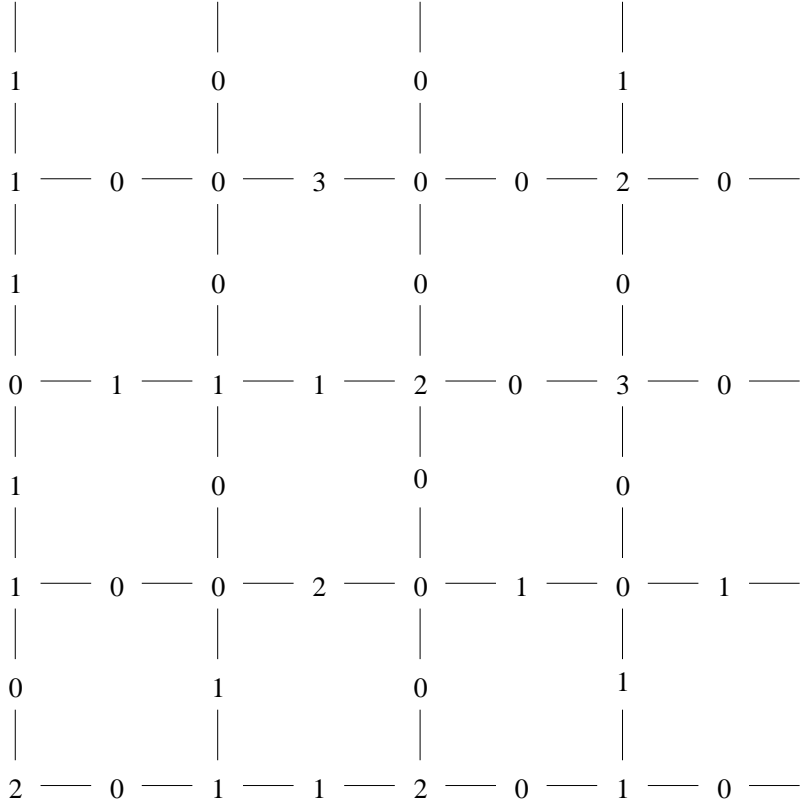}
\caption{ \label{unconf} \em An example of $N=3$ monomer-dimer configuration 
on a $4\times 4$ lattice.}
\end{center}
\end{figure}

The partition function of the model is given by
\begin{equation}
\label{opfunc}
Z = \int \ \prod_{(x,\mu)}\ [dU_\mu(x)]\ 
\prod_x \ [d\psi(x)\ d\bar\psi(x)] \ \exp(- S),
\end{equation}
where $[dU]$ is the Haar measure on the $U(N)$ group,
$[d\psi(x) d\bar\psi(x)]$ represent Grassmann integration. 
The model with $t=1$ was considered earlier in \cite{Wol84}, where it was 
shown that the integrals in eq.(\ref{opfunc}) can be performed analytically
and the partition function can be rewritten as a sum over positive
definite Boltzmann weights associated to monomer-dimer configurations. 
Let us see this explicitly by setting $t=1$. First note that the integral over 
the gauge fields can be done one link at a time in the background of the 
Grassmann fields. In \cite{Wol84} it was shown that
\begin{eqnarray}
\int  \ [dU_\mu(x)] \ \ \exp\Big[\bar\psi(x) U_\mu(x) \psi(x+\hat{\mu}) 
 &-& \bar\psi(x+\hat{\mu})(U_\mu(x))^\dagger\psi(x)\Big]
\nonumber \\
= \ \ \sum_{b=0}^N \ \ \frac{(N-b)!}{N! b!}&&  
\left[\bar\psi(x)\psi(x)
\bar\psi(x+\hat{\mu})\psi(x+\hat{\mu})\right]^{b}
\end{eqnarray}
and
\begin{equation}
\int \ [d\psi(x)\ d \bar\psi(x)] \
\exp(m \bar\psi(x)\psi(x)) 
\left[\bar\psi(x)\psi(x)\right]^{(N-n)}
\ = \ \frac{N!}{n!} m^{n}.
\end{equation}
Using these relations we can rewrite the partition function as
\begin{equation}
Z \;=\; \sum_{[n,b]} \;\;\;\;\;
\prod_{x,\mu}\;\frac{(N-b_\mu(x))!}{b_\mu(x)! N!}\;\;\; 
\prod_x \frac{N!}{n_x!}\;m^{n_x}
\label{bpf}
\end{equation}
where  $n_x=0,1,...,N$ is the number of monomers located at the 
site $x$ and $b_\mu(x)=0,1,2,...,N$ is the number of dimers 
located on the bond connecting the sites $x$ and $x+\hat{\mu}$. 
The configurations $[n,b]$ denote the sets of monomer values $n=\{n_x\}$
on all sites and bond values $b=\{b_{\mu}(x)\}$ on all links which 
satisfy the constraint
\begin{equation}
\label{constraint}
n_x + \sum_\mu \; b_\mu(x) + b_{-\mu}(x) \ = \ N
\end{equation}
on each site $x$. We use the definition $b_{-\mu}(x) \equiv 
b_\mu(x-\hat{\mu})$. In other words the total number of monomers and 
dimers associated to a site must always be $N$. The only effect of a 
general $t$ is that every dimer along the $\mu=1$ link is weighted 
by an extra factor of $t^2$. For illustration, an $N=3$ monomer-dimer 
configuration is shown in figure \ref{unconf}. 

When $m=0$ the action given in eq. (\ref{fact}) is invariant 
under $U(1)$ chiral transformations,
\begin{eqnarray}
\psi(x) \rightarrow \mathrm{e}^{i\theta}\psi(x),&
\bar\psi(x) \rightarrow \bar\psi(x)\mathrm{e}^{i\theta},&
\mbox{for $x$ even} \nonumber \\
\psi(x) \rightarrow \mathrm{e}^{-i\theta}\psi(x),&
\bar\psi(x) \rightarrow \bar\psi(x)\mathrm{e}^{-i\theta},&
\mbox{for $x$ odd}.
\end{eqnarray}
A site $x$ is defined to be even (odd) if $x_1+x_2+...+x_d$ is 
even (odd). To study the properties of the vacuum under 
chiral transformations one usually defines the chiral condensate,
\begin{equation}
\langle \bar\psi\psi\rangle \ \equiv \
\frac{1}{V}\sum_{x} \langle \bar\psi\psi(x)\ \rangle  = 
\frac{1}{Z} 
\int \ [dU d\psi d\bar\psi] \left\{\frac{1}{V} 
\sum_x \bar\psi(x)\psi(x)\right\}\ \mathrm{e}^{- S}.
\label{cond}
\end{equation}
In a finite volume the vacuum is chirally symmetric in the chiral
limit. A consequence of this symmetry is that
\begin{equation}
\lim_{m\rightarrow 0} \langle \bar\psi\psi\rangle = 0.
\end{equation}
In the monomer-dimer model 
\begin{equation}
\langle \bar\psi\psi\rangle \ = \ \frac{1}{m V} \sum_x \langle n_x \rangle
\end{equation}
where $ \langle n_x \rangle$ is the average number of monomers on the
site $x$. In the chiral limit at finite volumes this average goes to 
zero as $m^2$, which again means the chiral condensate vanishes.
Mean field calculations on the other hand suggest that the strong coupling
vacuum breaks the chiral symmetry spontaneously \cite{Kaw81,Klu83}. This 
means
\begin{equation}
\lim_{m\rightarrow 0} \lim_{V\rightarrow \infty} 
\langle \bar\psi\psi\rangle \neq 0
\label{ivcc}
\end{equation}
We will call this non-vanishing quantity the infinite volume chiral condensate.
It would be interesting to calculate this condensate using the new algorithm.
However, using eq.(\ref{ivcc}) to extract it is difficult since one 
needs to add a small fermion mass term to the action and then perform a 
careful scaling analysis of the chiral condensate with respect to both 
the volume and the mass. Since our algorithm can work efficiently at $m=0$
in a finite volume we will instead calculate the chiral susceptibility, 
defined by
\begin{equation}
\chi \equiv 
\frac{1}{V}\sum_{x,y} \langle \bar\psi\psi(x)\ \bar\psi\psi(y) \rangle 
= \frac{1}{Z} 
\int \ [dU d\psi d\bar\psi] \left\{\frac{1}{V} 
\sum_{x,y}\ \bar\psi(x)\psi(x)\ \bar\psi(y)\psi(y)\right\}\ \mathrm{e}^{- S}.
\label{csus}
\end{equation}
It is easy to show that 
\begin{equation}
\chi = V \Big(\langle \bar\psi\psi \rangle\Big)^2 
+ \frac{\partial \langle \bar\psi\psi\rangle}{\partial m}
\end{equation}
Usually, the chiral susceptibility is defined as the first derivative 
of the chiral condensate with respect to the mass. However, our definition 
includes the disconnected term proportional to the square of the infinite 
volume chiral condensate. With this definition, in a cubical box 
$V \equiv L^d$, we can argue that
\begin{equation}
\chi = \left\{ \begin{array}{cl}
\mbox{const.} L^d  &
\mbox{in a phase with broken chiral symmetry} \cr
 \mbox{const.} L^{2-\eta} &
\mbox{at a chirally symmetric critical point} \cr
 \mbox{const}. &
\mbox{in the chirally symmetric massive phase} 
\end{array} \right.
\label{sus}
\end{equation}
in the large volume limit. In the phase with broken chiral symmetry
the coefficient of $L^d$ is the square of the infinite volume chiral 
condensate which can be extracted by studying the large volume behavior 
of $\chi$ and fitting the data to the above form.
With conventional algorithms $\chi$ is very difficult to compute since 
it is a noisy observable. Our new approach on the other hand allows us 
to measure it very accurately even in large volumes. Although we have
written eq.(\ref{sus}) for a general $d$, the Mermin-Wagner-Coleman 
theorem \cite{Mer66,Col73} forbids a phase with broken chiral symmetry 
in two dimensions. In that case only the critical or the massive phases
are possible.

\section{DIRECTED PATH ALGORITHM}
\label{algo}

The constraints imposed by eq.(\ref{constraint}) make it difficult
to design algorithms for the monomer-dimer systems near the chiral 
limit. When $m=0$, configurations with monomers have vanishing weight 
in the partition function and local algorithms find it difficult to 
update the remaining constrained dimer configurations efficiently.
Cluster algorithms on the other hand can deal with constraints very 
efficiently. An analogue of this problem in a well-known setting 
is the following: Consider for example a quantum spin-half 
system where the number of 
``up'' spins and ``down'' spins are individually conserved. A typical
configuration of such a system is represented by a world-line of spins. 
While local algorithms find it extremely difficult to update such
configurations, the loop cluster algorithm is very efficient 
\cite{Eve93}. In \cite{Cha01,Cha02} we discovered that the monomer-dimer 
model can also be rewritten in terms of loop variables and found them to 
be convenient tools to update the system when $m=0$. Unfortunately, the 
Metropolis algorithm that was designed to update the loops turns out to 
be inefficient. Certain large clusters can only be updated with small 
probabilities and the algorithm develops long autocorrelation times as 
the lattice size grows. This problem is similar to the problem encountered 
in an anti-ferromagnetic quantum spin-half model in the presence of a 
uniform magnetic field. The loop cluster algorithm which works efficiently 
in updating the model in the absence of magnetic fields, becomes 
exponentially inefficient in their presence. Again, some large clusters 
get frozen and cannot be updated. 

Recently, a new algorithm called the ``directed loop'' algorithm 
was proposed for the antiferromagnetic model in a magnetic field 
\cite{San02}. This algorithm is extremely efficient even for large magnetic 
fields. Here we extend this algorithm to study strong coupling gauge theories.
In this section we construct a ``directed path cluster algorithm'' to update 
the monomer-dimer configurations. Our construction is such that the algorithm 
does not change the number of monomers in a given configuration. Hence, 
it is only ergodic in the chiral limit where the number of monomers is
strictly zero and does not change. When $m\neq 0$ it is necessary to 
supplement it with an update that changes the number of monomers. For
this we can use any local algorithm. We will briefly review one such 
algorithm at the end of this section for completeness.

The basic idea behind our algorithm is to create a monomer at a site.
However, in order to do this while satisfying the constraints it is
necessary to remove a dimer from one of the links connected to the site and
create another monomer on the other end of the bond. This neighboring
monomer is then moved along a  directed path, while satisfying detailed
balance at each step until it encounters a partner such that both 
monomers can be removed by creating a dimer. Thus, at the end of a directed 
path update, the number of monomers remains fixed. It is interesting to 
note that at every stage, between the start and the end of the path, we 
sample configurations that are relevant in the computation of two point 
correlation function of monomers. Hence, these intermediate configurations 
can be used to measure certain observables. We will discuss this feature 
of the algorithm in the next section.

In order to explain the algorithm better and show that it obeys detailed
balance it is useful to develop some notation. We begin by dividing the 
sites of the lattice into {\em active} and {\em passive} sites. If the 
first site we pick during the update is even then all even sites are 
defined to be active and all odd sites become passive. On the other 
hand if the first site happens to be odd, then all odd sites become 
active and even sites become passive. Each active (passive) site is 
associated with an active (passive) block which includes the site 
with the information $n$ about number of monomers on it and the $2d$ bonds 
connected to the site with the information $b_1,b_{-1},...b_d,b_{-d}$ of 
their dimer content. Due to the constraint eq.(\ref{constraint}) we must 
have
\begin{equation}
n + b_1 + b_{-1}+ + ...+ b_d + b_{-d} = N.
\end{equation}
We will represent the block symbolically by $(n,b_1,b_{-1},...,b_d,b_{-d})$.
Its Boltzmann weight is defined to be
\begin{equation}
W_{\rm active} = \frac{N!}{n!} \prod_{\mu=1}^d 
\frac{(N-b_\mu)!(N-b_{-\mu})!}{b_\mu! b_{-\mu}! (N!)^2}
\label{actw}
\end{equation}
if it is an active block and 
\begin{equation}
W_{\rm passive} = \frac{N!}{n!}\ (t^2)^{(b_1 + b_{-1})}
\label{pasw}
\end{equation}
if it is a passive block. We have dropped factors of the mass because
our objective is not to change the total number of monomers in the
configuration. It is easy to verify that the product of
the weights of all the active and passive blocks in a configuration 
is equal to the Boltzmann weight of the configuration up to some power
of the mass.

\begin{figure}[htb]
\vskip0.2in
\begin{center}
\includegraphics[width=25pc]{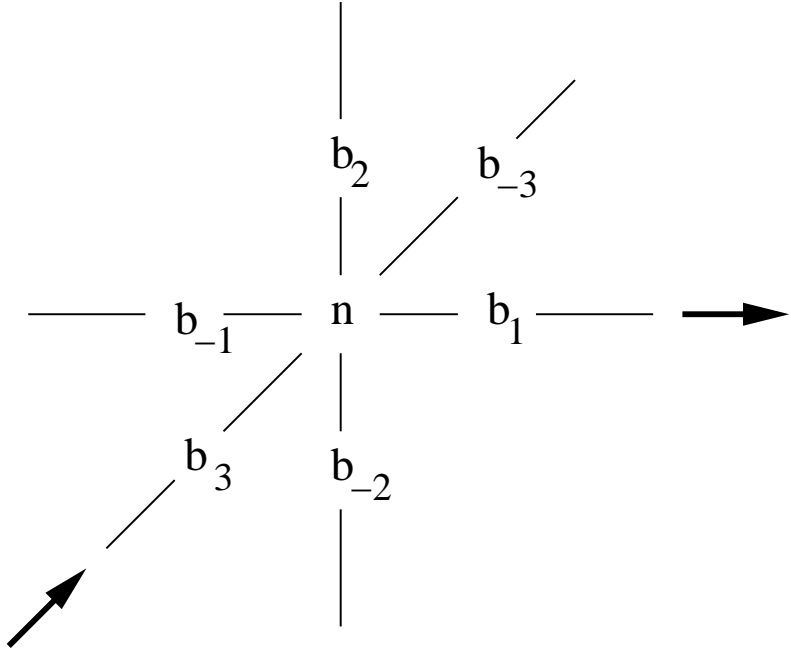}
\caption{ \label{block} \em A typical directed path passing through
a 3d block. The path enters the block through the $\mu=3$ bond and leaves
through the $\mu=1$ bond. The block and the path are represented by
$(n,b_1\odot,b_{-1},b_2,b_{-2},b_3\otimes,b_{-3})$.}
\end{center}
\vskip-0.2in
\end{figure}

A directed path update begins by entering a site at random. By 
definition that site belongs to an active block and the path enters
the block through the site. Given an in-coming path the algorithm
works on finding an outgoing path with a probability that satisfies
detailed balance. The out-going path can either be one of the $2d$ 
bonds or the starting site itself. As we will see shortly, the 
correct probability to choose the outgoing path to be the starting
site is
\begin{equation}
\label{p1}
P_{ss} = \frac{n}{N}
\end{equation}
and the probability to choose the bond in the $\mu = 1,-1,2,-2,..,d,-d$ 
direction is
\begin{equation}
\label{p2}
P_{s\mu} = \frac{b_\mu}{N}.
\end{equation}
If the out-going path is chosen to be the starting site then the directed path 
update already ends. The configuration is returned without being updated. 
Otherwise the path goes out through the chosen bond. As it leaves the 
active block it updates it by adding a monomer on the incoming site and 
then removing the monomer again if the path terminates immediately, or 
reducing the bond number on the outgoing link if the path continues.
Let us invent a notation to describe this process.
We use the symbol $\otimes$ next to a bond or a site to indicate 
in-coming path into the block and use $\odot$ to indicate the out-going 
path. For example $(n\otimes,b_1,b_{-1}\odot,...,b_d,b_{-d})$ means 
that the path came into the block through the site and left the block 
through the $\mu=-1$ bond.  When the path leaves the (active) block, 
the updated block is given by $(n+1,b_1,b_{-1}-1,...,b_d,b_{-d})$.
Figure \ref{block} shows a typical directed path through a three 
dimensional block.

When the path leaves an active block through a bond it enters the
neighboring passive block through the same bond. It is then forced to
leave through one of the remaining $2d-1$ bonds with probabilities given 
below. Since now the value of $t^2$ is important, we distinguish two cases:
$t^2 \leq 1$ and $t^2 \geq 1$. If the incoming path is through the 
$\mu= 1$ ($\mu = -1$) bond, 
the outgoing path can either be the $\mu= -1$ ($\mu=1$) bond or any one of 
$\mu \neq \pm 1$ bonds.  The probability to pick $\mu = -1$ ($\mu = 1$) is 
given by
\begin{equation}
P_{1,-1} = P_{-1,1} = \left\{ \begin{array}{cc}
  \frac{t^2}{2d - 2 + t^2} & t^2 \leq 1 \cr \cr
  \frac{2t^2 - 1}{2d - 3 + 2t^2} & t^2 \geq 1 \end{array} \right.,
\label{p3}
\end{equation}
while the probability to choose any one of the $\mu \neq \pm 1$ is given by
\begin{equation}
P_{1,\mu} = P_{-1,\mu} = \left\{ \begin{array}{cc}
  \frac{1}{2d - 2 + t^2} & t^2 \leq 1 \cr \cr
  \frac{1}{2d - 3 + 2t^2} & t^2 \geq 1 \end{array} \right..
\label{p4}
\end{equation}
However, if the path comes into the block through one of the 
$\mu \neq \pm 1$ bonds, then the outgoing path is chosen to be 
the $\mu = 1$ or $\mu = -1$ bond with probability
\begin{equation}
P_{\mu,1} = P_{\mu,-1} = \left\{ \begin{array}{cc}
  \frac{t^2}{2d - 2 + t^2} & t^2 \leq 1 \cr \cr
  \frac{t^2}{2d - 3 + 2t^2} & t^2 \geq 1 \end{array} \right.
\label{p5}
\end{equation}
and is chosen to be one of the $\nu \neq \pm 1,\mu$ bonds with probability
\begin{equation}
P_{\mu,\nu} = \left\{ \begin{array}{cc}
  \frac{2d-2-t^2}{(2d - 2 + t^2)(2d-3)} & t^2 \leq 1 \cr \cr
  \frac{1}{2d - 3 + 2t^2} & t^2 \geq 1 \end{array} \right.
\label{p6}
\end{equation}
As the path leaves a passive block, the block itself is
updated by lowering the dimer number on the incoming bond and 
raising the dimer number on the outgoing bond. 
We note that in eqs.(\ref{p3}-\ref{p6}) the probability definitions 
for $t\leq 1$ and $t\geq 1$ can be extended to the range $t^2 < 2d-2$ 
and $t^2 > 1/2$, respectively. This is a reflection of the fact that 
the requirement of detailed balance does not determine all the 
probabilities uniquely.

When the path leaves the passive block it enters the adjacent active 
block through an in-coming bond. In this case it can now leave through 
one of the remaining $2d-1$ bonds or the site. Given the 
in-coming path to be through the bond $\nu$, the probability to choose the 
bond $\mu$ as the out-going path is 
\begin{equation}
\label{p7}
P_{\nu\mu} = \frac{b_\mu}{N-b_\nu}
\end{equation}
and the probability to choose the site as the outgoing path is
\begin{equation}
\label{p8}
P_{\nu s} = \frac{n}{N-b_\nu}.
\end{equation}
If the path leaves through a bond then it continues into another passive 
block and goes forward as already discussed above. However, if it exits 
through the site then the path ends. In either case the active block is 
again updated by increasing the dimer content of the incoming bond by 
one and decreasing the dimer content of the outgoing bond or decreasing 
the monomer content of the outgoing site by one. 

Thus, we see that the algorithm always enters and exits through the sites
of active blocks. It increases the monomer number on the incoming site by 
one and decreases the monomer number at the exiting site by one, and thus 
preserving the total number of monomers. Of course, the site at which the 
directed path enters may or may not be the site in which it leaves, although 
for $m=0$ these two sites are forced to be the same since the weight of a 
configuration which contains monomers vanishes. Thus, in the chiral limit, 
the directed path is always a closed loop. An important difference between 
active and passive blocks is that the paths always enter and leave a
passive block through a bond. Hence the monomer content of sites belonging
to passive blocks never change during a single directed path update.
The paths always enter a passive block on a bond which has at least one 
dimer on it. 

It remains to be shown that eqs.(\ref{p1}-\ref{p8}) satisfy detailed 
balance with respect to the Boltzmann weights given in 
eqs.~(\ref{actw}) and (\ref{pasw}). 
First consider the active block represented by 
$(n,b_1,b_{-1},..,b_\mu,..,b_\nu\otimes,...b_{d},b_{-d})$
with the in-coming path specified to be the bond $\nu$. The probability to 
choose the bond $\mu$ as the out-going bond is given by eq.(\ref{p7}). 
After the update the active block is changed to
$(n,b_1,b_{-1},..,b_\mu-1,..,b_\nu+1,...b_{d},b_{-d})$. In 
order to prove detailed balance one is interested in the reverse process. For
this, one should start with the configuration
$(n,b_1,b_{-1},..,b_\mu-1\otimes,..,b_\nu+1,...b_{d},b_{-d})$,
where now the in-coming path is the bond $\mu$, and figure out the
probability of choosing $\nu$ as the out-going path. This turns out
to be
\begin{equation}
P'_{\mu\nu} = \frac{b_\nu+1}{N-b_\mu+1}
\end{equation}
We see that these probabilities and the weights given in eq.(\ref{actw}) 
satisfy detailed balance:
\begin{equation}
\label{db1}
\frac{b_\mu}{(N-b_\nu)}
\frac{(N-b_\mu)!}{b_\mu!}\frac{(N-b_\nu)!}{b_\nu!}
= \frac{b_\nu+1}{(N-b_\mu+1)} 
 \frac{(N-b_\mu+1)!}{(b_\mu-1)!}\frac{(N-b_\nu-1)!}{(b_\nu+1)!}.
\end{equation}
In the above detailed balance equation we have only considered factors of 
the Boltzmann weight that change during the update. 

In the case of the passive block it is again easy to show detailed balance.
For example consider the case when the incoming path is through the $\nu=1$ 
bond and the outgoing bond is through the $\mu \neq \pm 1$ bond. We can 
represent this path by 
$(n,b_1\otimes,b_{-1},..,b_\mu\odot,...b_{d},b_{-d})$. After the update 
the new weight of the block is equal to the old weight divided by $t^2$.
The reverse process, starting from the updated block looks like
$(n,b_1-1\odot,b_{-1},..,b_\mu+1\otimes,...b_{d},b_{-d})$. The detailed
balance requires
\begin{equation}
P_{1,\mu} t^2 = P_{\mu,1}
\end{equation}
which is indeed true when we use eqs.(\ref{p4},\ref{p5}). Using a similar
method one can show that other choices of incoming and outgoing paths on 
both active and passive blocks obey detailed balance locally at each 
update. There is one exception which we discuss below.

Consider the case when the in-coming path is a 
site and the out-going path is a bond and its reverse process. We represent 
this case by $(n\otimes,b_1,b_{-1},..,b_\mu\odot,..,b_\nu,...b_{d},b_{-d})$, 
for which the probability is given by eq.(\ref{p2}). The reverse process 
on the other hand can be represented by
$(n+1\odot,b_1,b_{-1},..,(b_\mu-1)\otimes,..,b_\nu,...b_{d},b_{-d})$, which has
the probability
\begin{equation}
P''_{\mu\nu} = \frac{n+1}{N-b_\mu+1}
\end{equation}
We now see that the detailed balance is not quite satisfied locally, since
\begin{equation}
\label{db2}
\frac{b_\mu}{N}
\frac{(N-b_\mu)!}{b_\mu!}\frac{N!}{n!}
\neq \frac{n+1}{(N-b_\mu+1)} 
 \frac{(N-b_\mu+1)!}{(b_\mu-1)!}\frac{N!}{(n+1)!}
\end{equation}
There is a extra factor $1/N$ that remains uncanceled while we go from the 
site to the bond. However, since the complete ``directed path'' update 
starts and ends on an active site, this extra factor must appear in each 
direction after a complete path update. This then guarantees that the 
complete forward and reverse directed paths indeed satisfy detailed balance.

Before we end this section let us briefly comment on the ergodicity of the
algorithm. When $m=0$ the ``directed path'' update is ergodic by itself. 
In order to see this, let us assume that $N=1$. In this case, any 
configuration can be transformed into any other configuration by  
a series of disconnected directed loop updates. The loops themselves can 
be identified by superimposing one configuration over the other. All 
dimers that differ in the two configurations connect the sites into 
disconnected loops. Clearly, there is a non-zero probability for
performing this specific series of directed loop updates and hence
to go from any configuration to any other. For $N>1$, 
one can easily extend this argument and prove ergodicity.

Since the directed path algorithm does not change the number of monomers, we 
need to supplement it with another algorithm that can change the number 
of monomers in a configuration when $m$ is nonzero, in order to make the 
algorithm ergodic. Many options are available. One can choose
a local algorithm based on either a Metropolis or a Heat Bath update. In 
the case of a Metropolis update for example, a simple algorithm would be
to pick a bond at random and propose to either break the dimer into two 
monomers or create a dimer from two monomers. This proposal is then accepted 
with a Metropolis accept reject step. We find that this combined algorithm
is reasonably efficient. We are currently exploring a more natural extension 
of the directed path algorithm to the massive case which avoids the 
additional local Metropolis step. This will be published elsewhere.

\section{OBSERVABLES}
\label{obs}

Let us for the moment assume that the directed path algorithm discussed
above can efficiently sample configurations that contribute to the partition 
function. If these configurations also contribute to an observable, 
then we may be able to compute the average of the observable efficiently.
However, for this to be true, the observable must get all the
contribution only from the ensemble of configurations sampled by
the algorithm and the value of the observable must not fluctuate 
much in this ensemble. Observables which satisfy these properties
will be referred to as ``diagonal'' observables. For example observables
like dimer-dimer correlations and spatial winding numbers associated 
to the global $U(1)$ chiral symmetry are diagonal observables. 
When $m$ is not very small, monomer correlations can also be treated
as diagonal observables. On the other hand when $m=0$, observables involving 
monomers do not fall in this category;
such observables get contributions from configurations with monomers, 
while the algorithm only produces zero-monomer configurations. In 
other words the configurations that are important 
to the partition function are not useful to measure observables. 
Such observables are difficult to
compute and will be referred to as ``non-diagonal'' observables. 
The chiral condensate and the chiral susceptibility are two
interesting examples of such observables. We will argue below that 
the directed path algorithm offers an efficient method to compute them.

From the discussion in the previous section we know that a directed 
path update starts on an active site, goes through an alternate 
series of both passive and active sites and ends on an active site. 
It is important to recognize that between the start and the end of 
each update we sample a configuration with exactly two additional 
monomers every time we visit a passive site; one monomer is located
at the starting active site and the other at the visited passive 
site. These intermediate configurations can be used to compute monomer 
correlations. To write down explicit relations we introduce an integer 
function $I(x,y)$ for each directed path update where $x,y$ are 
lattice sites. This function is defined as follows: Before we start 
the directed path update we set $I(x,y)=0$ for all values of $x,y$; 
we then add one to $I(x,y)$ if the directed path update starts on the 
active site $y$ and every time it visits the passive site $x$. Now 
consider a monomer-dimer configuration with $n_i$ monomers located 
at $y_i,i=1,2,...,k$. For this configuration we can define a one 
point function and a two point function as follows:
\begin{eqnarray}
S_1(x) &=& \frac{m N}{2d-2+2t^2}\ \sum_{z} \ I(x,z)
\label{corrobs1}
\\
S_2(x,y) &=& \frac{N}{2d-2+2t^2}
\left[ \ I(x,y) + \left( \sum_{i=1}^k n_i \delta_{y,y_i} \right)\
\sum_z \ I(x,z) \right]
\label{corrobs2}
\end{eqnarray}
It is possible to show that
\begin{equation}
\frac{1}{V}\langle \ \bar\psi\psi(x)\ \rangle
\;=\; \Big \langle S_1(x) \ \Big \rangle
\label{optcorr}
\end{equation}
\begin{equation}
\frac{1}{V}\langle \ \bar\psi\psi(x)\  \bar\psi\psi(y)\ \rangle
\;=\; \Big \langle \ S_2(x,y) \ \Big \rangle.
\label{tptcorr}
\end{equation}
In appendix \ref{obsder} we give a brief derivation of these results. 
Using eqs. (\ref{optcorr}) and (\ref{tptcorr}) one finds
\begin{equation}
\langle \bar\psi\psi \rangle =
\label{cc} \Big\langle \sum_{x} \bar{S}_1(x) \Big \rangle
\end{equation}
and
\begin{equation}
\label{susc}
\chi = \Big\langle (n + 1) \sum_{x,y} \bar{S}_2(x,y) \Big\rangle,
\end{equation}
where we have used $n = n_1+n_2+...+n_k$ to denote the total number of
monomers in the configuration. In eqs. (\ref{optcorr}-\ref{susc}) the 
average on the right hand side is taken over the ensemble of configurations 
generated in the directed path algorithm. 

As the derivation in the appendix shows, some additional work is necessary
to derive expressions for the non-diagonal observables in the directed 
path algorithm. However, it is usually possible to find expressions for
most observables. In fact similar expressions have been obtained in the 
context of other cluster algorithms \cite{Bro98,Cha00}. 
In the next section we will demonstrate the correctness of the relations
(\ref{cc}) and (\ref{susc}) by comparing the results obtained using them 
with exact calculations to a high accuracy. 

It is clear that one can also compute many other interesting quantities.
All observables involving two monomers can be obtained using 
eq.(\ref{tptcorr}).  Higher point monomer 
correlations functions can also be computed but need additional 
work. They will be discussed elsewhere.

\section{TESTING THE ALGORITHM}

We have tested the directed path algorithm and the expressions of
the chiral condensate (\ref{cc}) and the susceptibility (\ref{susc})
in various dimensions for small lattices where exact calculations are 
possible. In this section we briefly review our tests. For a finite 
lattice with $V$ sites, the partition function (eq.(\ref{bpf})) is an 
even polynomial in the mass
\begin{equation}
Z(m)=c_0+c_2m^2+c_4m^4+\dots+c_{NV}m^{NV}.
\label{e1}
\end{equation}
The coefficients $c_{2n}$ are functions of $t^2$, although we suppress 
this dependence in the notation. The absence of terms with odd powers 
of $m$ in eq.(\ref{e1}) is a consequence of the remnant $U(1)$ chiral 
symmetry. Every configuration $[n,b]$ can only have an even number 
of monomers. $NV$ is the maximum number of monomers allowed.

The condensate and the susceptibility can be obtained from the partition 
function using the relations
\begin{equation}
\la\bpsi\psi\ra=\frac{1}{V}\,\frac{1}{Z(m)}\,\frac{dZ(m)}{dm}\qquad,
\quad\chi=\frac{1}{V}\,\frac{1}{Z(m)}\,
\frac{d^2 Z(m)}{dm^2}\,
\label{e2}
\end{equation}
Thus, once the coefficients $c_0,c_2,\dots,c_{NV}$ are known these quantities 
can be determined. In the appendix we give expressions for these 
coefficients in some simple cases. When $m=0$ the 
condensate vanishes as it should and $\chi = 2 c_2/ V c_0$. Hence,in
some cases where calculating all the coefficients is difficult, we
give only the coefficients $c_0$ and $c_2$. 

In tables \ref{exactpsi} and \ref{exactsus} we compare the condensate 
and susceptibility obtained using (\ref{cc}) and (\ref{susc}) with 
exact results.
\begin{table}[htb]
\vskip-0.2in
\begin{center}
\caption{ {\label{exactpsi} Chiral Condensate: Algorithm vs. Exact results.}}
\vskip0.1in
\begin{tabular}{|c|c|c|c|l|c|}
\multicolumn{6}{c}{chiral condensate} \\ \hline
$N$ & Lattice Size & $m$ & $t^2$ &  Exact & Algorithm \\
\hline
1 & $2\times 2$ & 0.3 & 1.3 & 0.12133... & 0.12135(4) \\ 
1 & $2\times 2 \times 2$ & 0.47 & 0.43 & 0.716859... & 0.71658(23) \\ 
2 & $2\times 2 \times 2$ & 0.1 & 3.6 & 0.067350... &  0.06736(2)\\ 
1 & $2\times 2 \times 4$ & 0.2 & 6.4 & 0.0206279... &  0.02063(3)\\ 
1 & $4\times 2 \times 2$ & 0.4 & 4.7 & 0.1263817... &  0.12643(5)\\ 
\hline
\end{tabular}
\end{center}
\end{table}

\begin{table}[htb]
\vskip-0.2in
\begin{center}
\caption{ {\label{exactsus} Chiral Susceptibility: 
Algorithm vs. Exact results.}}
\vskip0.1in
\begin{tabular}{|c|c|c|c|l|c|}
\multicolumn{6}{c}{chiral susceptibility} \\ \hline
$N$ & Lattice Size & $m$ & $t^2$ &  Exact & Algorithm \\
\hline
1 & $8 \times 8$ & 0.0 & 1.0 &   5.27221... &  5.2722(2) \\
3 & $4 \times 4$ & 0.0 & 1.0 &  14.1595... & 14.159(8) \\
30 & $2 \times 2$ & 0.0 & 1.0 &  338.534... & 338.2(8) \\
1 & $2\times 2 \times 2$ & 0.13 & 1.2 & 0.57028... & 0.5702(2) \\ 
3 & $2\times 2 \times 2$ & 0.0 &  1.4 & 4.38869... & 4.3884(13) \\ 
1 & $2\times 2 \times 4$ & 0.0 &  3.2 & 0.25766... & 0.2576(1) \\ 
2 & $2\times 2 \times 4$ & 0.0 &  1.4 &  3.17378... & 3.1741(9) \\ 
1 & $4\times 2 \times 2$ & 0.0 &  0.37 &  0.69463... & 0.6945(3) \\ 
2 & $4\times 2 \times 2$ & 0.0 &  3.2 &  2.24205... & 2.2418(5). \\ 
1 & $2\times 2 \times 2 \times 2$ & 0.0 &  5.3 &  0.15401... & 0.15397(6) \\ 
1 & $2\times 2 \times 2 \times 2$ & 0.0 &  1.2 &  0.80231... & 0.8022(2) \\ 
\hline
\end{tabular}
\end{center}
\end{table}
In order to check if the algorithm performs well for large lattices, we 
have derived an exact expression for $\chi$ on a $2 \times L$ lattice 
when $m=0$, $N=1$, $t=1$. After a lengthy calculation we found 
\begin{equation}
\chi=\frac{{\textstyle \frac{1}{2\sqrt{2}}}((\sqrt{2}+1)^{L+1}
+(\sqrt{2}-1)^{L+1})-1}
{(\sqrt{2}+1)^L+(\sqrt{2}-1)^L+2}
\label{e12}
\end{equation}
For $L\rightarrow \infty$ this becomes
\begin{equation}
\lim_{L\to\infty}\;\chi=\frac{(\sqrt{2}+1)}{2\sqrt{2}}=0.8535534...
\label{e13}
\end{equation}
Our algorithm yields $\chi=0.8535(3)$ when $L=1024$.

\section{PERFORMANCE OF THE ALGORITHM}
\label{perf}

An important feature of a good algorithm is a short auto-correlation 
time which we denote as $\tau$. Due to critical slowing down the 
auto-correlation time increases with increasing correlation lengths. 
One expects $\tau \propto \xi^z$ where $\xi$ is the correlation length
and $z$ is the dynamical critical exponent of the algorithm.
For most local algorithms $1 \leq z \leq 2$. Many efficient cluster 
algorithms on the other hand are known to have $0 < z \leq 1$. In this 
section we estimate $z$ for the $N=1$ algorithm in two dimensions
for $m=0$ and $t^2=1$.

Although the auto-correlation time depends only on the algorithm, a 
more useful quantity in practice is the integrated auto-correlation 
time $\tau_{\rm int}$ defined for a given observable using the relation
\begin{equation}
\tau_{int} = \frac{1}{2}\sum_{t=0}^{t_{\rm max}} G(t),
\end{equation}
where
\begin{equation}
 G(t) =
\frac{\langle [{\cal O}_{i+t} - \langle {\cal O}\rangle] 
[{\cal O}_i - \langle {\cal O}\rangle] \rangle}
{\langle [{\cal O}_i - \langle {\cal O}\rangle] 
[{\cal O}_i - \langle {\cal O}\rangle] \rangle}
\label{tint}
\end{equation}
with ${\cal O}_i$ being the value of the observable generated by the 
algorithm in the $i^{\rm th}$ sweep and $\langle {\cal O}\rangle$
refers to its average. Ideally when the correlation function $G(t)$ 
is known exactly then $t_{\rm max}$ can be set to infinity. However,
in practice due to a finite data sample, $G(t)$ is determined with 
growing relative errors as $t$ increases. Hence, $t_{\rm max}$ must be 
chosen carefully \cite{Wol89}. It is also important to normalize 
$\tau_{\rm int}$ such that the definition of a sweep does not enter 
into it. Here, every directed loop update after a site is picked at 
random is defined as a sweep. This means that on an average it takes 
as many as $f = L^2/\langle {\cal N}_{\rm passive} \rangle$ sweeps
( ${\cal N}_{\rm passive}$ is the number of passive sites encountered
in the directed path update) before we update
every degree of freedom once. Hence we divide the answer obtained from 
eq. (\ref{tint}) by $f$ and define that as the integrated auto-correlation
time.

\begin{figure}[htb]
\vskip0.3in
\begin{center}
\includegraphics[width=25pc]{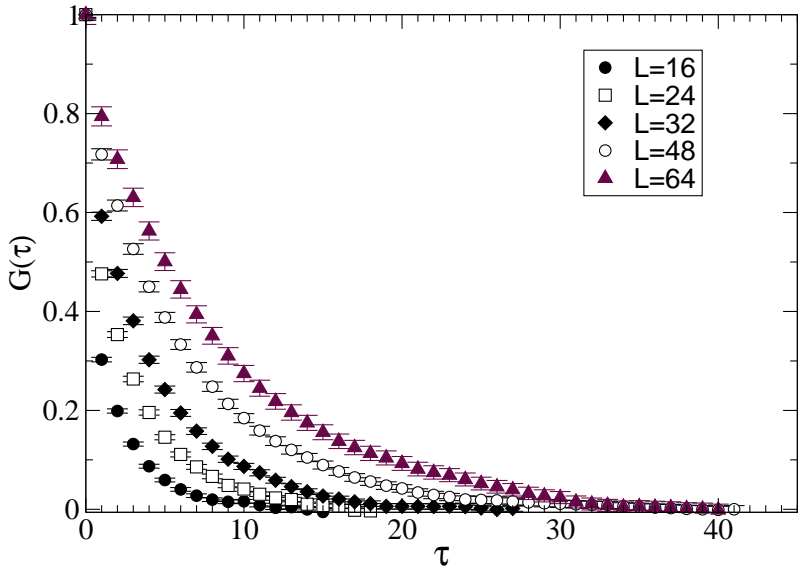}
\vskip0.1in
\caption{ \label{corr} \em The auto-correlation function $G(t)$ is shown
as a function of $t$ for $L=16,24,32,48$ and $64$}
\end{center}
\end{figure}

\begin{figure}[htb]
\vskip0.3in
\begin{center}
\includegraphics[width=25pc]{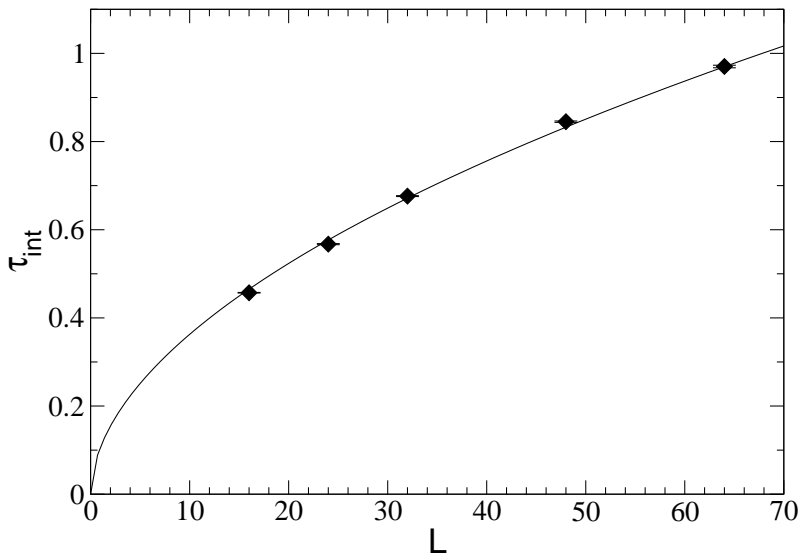}
\caption{ \label{tintfig} \em Integrated auto-correlation time as a function
of $L$.}
\end{center}
\vskip-0.2in
\end{figure}

Here we estimate $\tau_{\rm int}$ for the chiral susceptibility. As we will
see in the next section, $\xi$ is infinite for the parameters we use
($N=1,d=2,m=0$), since there are massless excitations. Hence, we study 
the behavior of $\tau_{\rm int}$ as a function of the lattice size $L$. 
Figure \ref{corr} plots the function 
$G(t)$ for $L=16,24,32,48$ and $64$. This graph suggests a simple choice 
for $t_{\rm max}$ for various $L$'s. We choose $t_{\rm max} = 12,16,20,30,30$ 
for the five values of increasing $L$. This defines $\tau_{\rm int}$ 
uniquely. We plot the results in figure \ref{tintfig}. The function 
$0.107 L^{0.53}$ (solid line) roughly captures the behavior of 
$\tau_{\rm int}$ as a function of $L$ suggesting that the dynamical 
critical exponent $z$ of our algorithm is around $0.5$. 

\begin{figure}[htb]
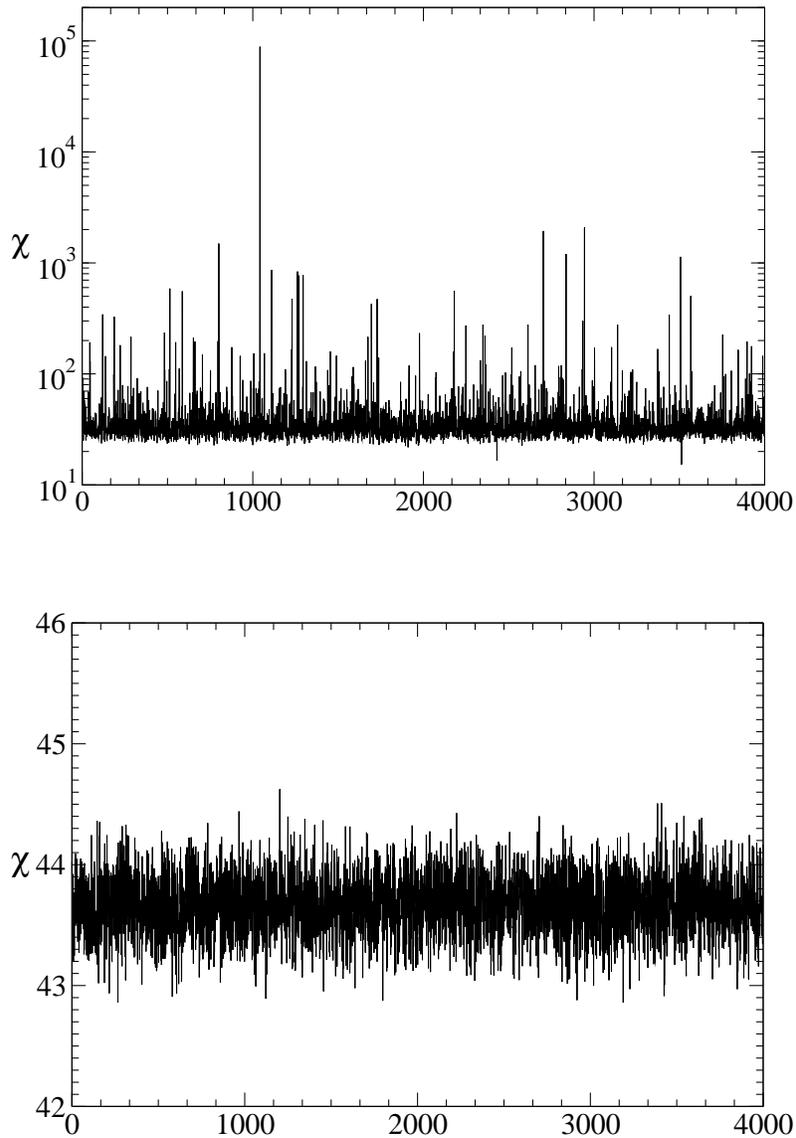

\vskip0.5in
\begin{center}
\vbox{
\includegraphics[width=25pc]{old.eps}
\vskip0.5in
\includegraphics[width=25pc]{new.eps}
\vskip0.2in\caption{ \label{algcmp} \em Comparison of simulation time 
histories of the chiral susceptibility for 4000 consecutive measurements 
between the old (top graph) and the new (bottom graph) algorithms for $L=32$.}
}
\end{center}
\vskip-0.2in
\end{figure}

Let us now compare the directed-path algorithm discussed here with the 
Metropolis algorithm developed in a previous work \cite{Cha01,Cha02}. 
This is of
interest since the results from the previous algorithm were quite 
unexpected and may be wrong. In the earlier work the two dimensional 
partition function was rewritten in terms of loop variables and the 
dimers along the loop were updated using a Metropolis accept-reject step. 
It was shown that the algorithm reproduced exact results with good
precision on small lattices. Based on the evidence from the directed 
path algorithm, we now argue that the auto-correlation time of the 
previous algorithm grows uncontrollably for large lattices. In figure 
\ref{algcmp} we compare $4000$ consecutive measurements of the
chiral susceptibility in the simulation time history at $m=0$ on a 
$32\times 32$ lattice, between the old and the new algorithms. 
In both cases, measurements are 
performed after a comparable amount of computer time. The fluctuations in
the old algorithm are clearly much larger. The average of $300,000$ 
measurements with the old algorithm gives a value of $41.8(6)$ as compared 
to $43.669(5)$ obtained from just 4000 measurements with the new algorithm. 
This clearly demonstrates the efficiency of the new algorithm in addition to 
showing that the earlier algorithm underestimates the final answer. Note
that there is one measurement of the order of $10^5$ visible in 
the old algorithm time history. If we remove this measurement from 
our analysis we obtain an answer of $41.0(2)$ even with $4000$ measurements.
This is consistent with the value obtained by averaging over the entire 
time history. 
Without more statistics it would be impossible to say whether these spikes 
are isolated events or contribute an important fraction to the final answer. 
However, our new and more efficient algorithm shows us that the contributions 
from these isolated spikes are indeed important and can contribute up to 
five percent to the final answer on a $32\times 32$ lattice. Comparing the 
results obtained from the new and the old algorithms for different lattice 
sizes, we now believe that these large spikes contribute even a larger 
percentage to the final answer on larger lattices. Most likely the old
algorithm is not able to tunnel between important sectors of the 
configuration 
space and develops large auto-correlation times at large volumes. Due to 
insufficient statistics we think that the contributions from the spikes were 
not sampled properly in the previous work and hence the final answers were 
systematically lower. This led to wrong conclusions.

\section{RESULTS IN THE CHIRAL LIMIT}

In this section we present some results obtained using the algorithm 
in the chiral limit. We focus on the issue of chiral symmetry breaking 
in various dimensions. As discussed earlier, the finite size scaling 
of the chiral susceptibility is an ideal observable for investigating 
this issue. All the results given below are obtained with $t=1$. 

\begin{figure}[htb]
\vskip0.4in
\begin{center}
\includegraphics[width=25pc]{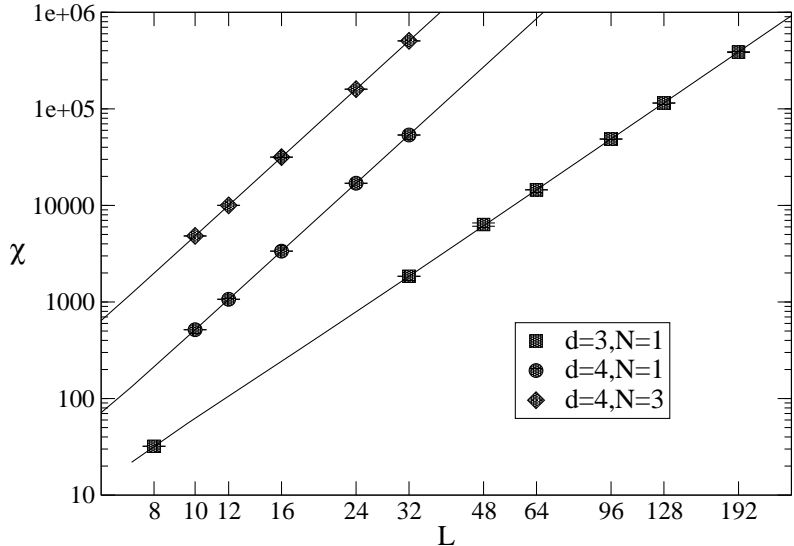}
\caption{ \label{susd} \em Finite size scaling of the Chiral 
Susceptibility in three and four dimensions. The solid lines 
represent fits of the data as described in the text.}
\end{center}
\vskip-0.2in
\end{figure}

Based on the work of \cite{Sal91} we know that at strong couplings 
chiral symmetry is broken in three and four dimensions. In figure \ref{susd} 
we plot the chiral susceptibility ($\chi$) as a function of the lattice 
size $L$ in $d=3,4$ when $N=1$ and in $d=4$ when $N=3$. For $d=4$
the data fits very well to the functional form $a L^d+b$, where $a$ and $b$
are constants. This functional form is motivated from chiral perturbation
theory. In the case of $d=3$ chiral perturbation theory suggests a form
$aL^d + bL^{4-d} + c$. Again the data fits well to this form over the
entire range. The values of all the fit parameters and the $\chi^2/$d.o.f. 
obtained from the fits are given in table \ref{tabd34}. The solid lines 
in figure \ref{susd} represent these fits.
\begin{table}[htb]
\begin{center}
\caption{\label{tabd34}}
\begin{tabular}{|c|c|c|c|c|c|}
\hline
 N & d & a & b & c & $\chi^2$/d.o.f. \\
\hline
1 & 3 & 0.05493(5) & 1.75(9) & -10(1) & 0.5 \\
1 & 4 & 0.05124(7) & 3.6(2) & -- & 0.5 \\
3 & 4 & 0.4823(2) & 12(5) & -- & 0.2 \\
\hline
\end{tabular}
\end{center}
\end{table}
The divergence of the susceptibility with $L$ is consistent with 
spontaneous breaking of chiral symmetry. Using eq.(\ref{sus}) 
we see that the infinite volume chiral condensate is given by
$\langle \bar\psi\psi\rangle = \sqrt{a}$. We find
\begin{equation}
\langle \bar\psi\psi\rangle = \left\{
\begin{array}{ccc} 
0.2343(2) & \mbox{for} & N=1, d=3 \cr
0.2264(2) & \mbox{for} & N=1, d=4 \cr
0.6945(3) & \mbox{for} & N=3, d=4.
\end{array}\right.
\end{equation}
We would like to remind the reader that the errors in the fitting
parameters do not include all the systematic errors that arise due 
to variations in the analysis procedures.

\begin{figure}[htb]
\vskip0.4in
\begin{center}
\includegraphics[width=25pc]{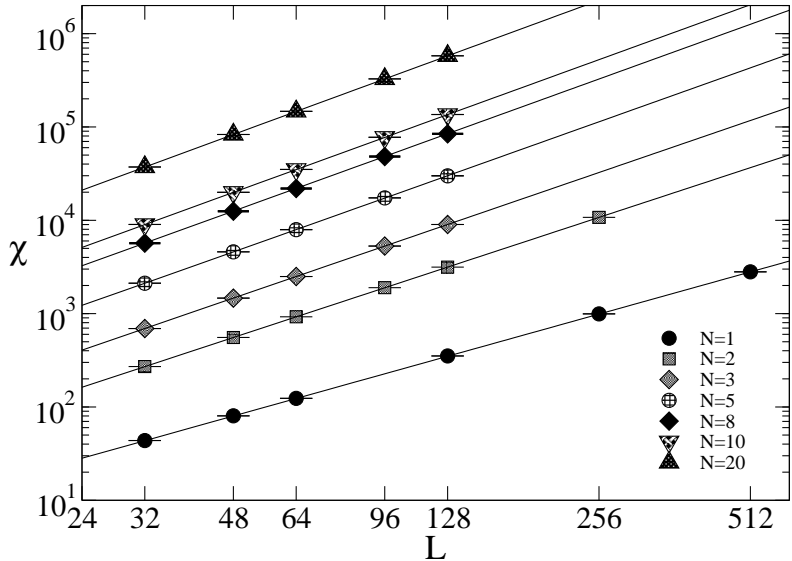}
\caption{ \label{sus2N} \em Finite size scaling of the Chiral 
Susceptibility in two dimensions for different values of $N$.
The solid lines are fits as discussed in the text.}
\end{center}
\vskip-0.2in
\end{figure}

In two dimensions the Mermin-Wagner-Coleman theorem forbids the formation of a
condensate \cite{Mer66,Col73}. However, it is well known that a theory 
with $U(1)$ symmetry can still contain massless excitations. Of course
there is always a possibility for the theory to be in the chirally symmetric
massive phase. To which phase does our model at $t=1$ belong?
The behavior of $\chi$ as a function of $L$ in both the possible phases
was already discussed in eq.(\ref{sus}). In figure \ref{sus2N} we plot 
$\chi$ as a function of $L$ for $N=1,2,3,5,8,10,20$. We find that all of 
the shown data fit reasonably well to the function $\chi = a L^{2-\eta}$.
The solid lines show these fits.
The values of $a,\eta$ and $\chi^2/$d.o.f. are given in table \ref{tabd2}. 
\begin{table}[htb]
\begin{center}
\caption{\label{tabd2}}
\begin{tabular}{|c|c|c|c|}
\hline
 $N$ & $a$ & $\eta$ & $\chi^2$/d.o.f. \\
\hline
 1 & 0.239(1) & 0.498(2) &  1.2 \\
 2 & 0.586(1) & 0.230(1) &  1.6 \\
 3 & 1.139(3) & 0.149(1) & 2.7 \\
 5 & 2.78(1) &  0.086(1) & 0.11 \\
 8 & 6.71(2) &  0.054(1) & 0.28 \\
 10 & 10.22(5) & 0.043(2) & 0.33 \\
 20 & 39.1(2) & 0.021(2) & 0.64 \\
\hline
\end{tabular}
\end{center}
\end{table}
This result strongly suggests that two dimensional staggered fermions
at strong couplings are in the critical massless phase.

The partition function of two dimensional dimer models ($N=1$, no monomers) 
on a planar lattice can be solved exactly \cite{Kas63,Tem61,Fis61}. In 
\cite{Fis63} a determinant formula for the two monomer correlation 
function ($\langle\bar\psi\psi(x)\ \bar\psi\psi(y)\rangle$) was derived 
in the infinite lattice setting. A combination of analytic and 
numerical analysis in \cite{Fis63} provided strong evidence that this
correlation function decays as $1/|x-y|^{1/2}$ when $|x-y|$ is large. 
This implies (in $N=1$ case) that the chiral susceptibility 
$\chi$  diverges as $L^{3/2}$ in large volumes, i.e. $\eta=0.5$. 
Our $N=1$ algorithm result 
 $\eta=0.498(2)$ is in excellent agreement with this semi-exact result.

\begin{figure}[htb]
\vskip0.4in
\begin{center}
\includegraphics[width=25pc]{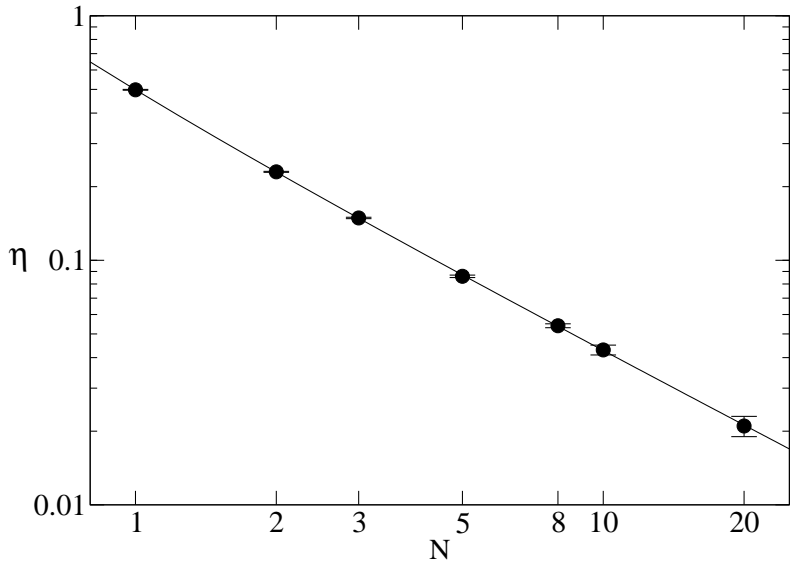}
\caption{ \label{eta} \em $\eta$ as a function of $N$. The solid line 
corresponds to the function $0.42/N + 0.078/N^2$.}
\end{center}
\vskip-0.2in
\end{figure}

At the other extreme, a large $N$ analysis leads to the 
conclusion that chiral symmetry is spontaneously broken even in two 
dimensions. Then the condensate must be non-zero. How does the 
condensate acquire a non-zero value? A resolution to this paradox 
was proposed by Witten in \cite{Wit78}. He predicted that 
$\eta \sim c/N$ for large $N$, so that when $N$ is strictly infinite 
$\eta=0$ and $\chi \sim L^2$ as expected in a phase with broken chiral 
symmetry in two dimensions (see eq.(\ref{sus})). Our data is consistent 
with this expectation. We find that all of our data fits well to the
form $\eta = 0.420(3)/N + 0.078(4)/N^2$  with $\chi^2/{\rm d.o.f.} \sim 0.33$.
Our data along with the fit is shown in figure \ref{eta}. The 
$1/N^2$ term in the fitting function is necessary to fit the data
for $N=1,2,3$.

\section{CONCLUSIONS}

In this article we have introduced a new type of cluster algorithm to 
update strongly coupled $U(N)$ lattice gauge theories with staggered
fermions in the chiral limit. We studied the performance of the algorithm
in two dimensions for a specific set of parameters and discovered that 
it has a dynamical critical exponent of roughly one half. We found the 
algorithm to be extremely efficient even in three and four dimensions. 
This progress allows us, for the first time, to study questions related 
to the chiral limit of a gauge theory beyond the mean field approximation.

As a first step we studied the question of chiral symmetry breaking. We
used finite size scaling of the chiral susceptibility as a tool
to address this question. We found clear evidence that the $U(1)$ chiral 
symmetry is spontaneously broken in three and four dimensions. On the
other hand, in two dimensions we found that the theory contains 
massless excitations although the chiral condensate was zero. This 
is consistent with the Mermin-Wagner-Coleman theorem. The critical 
exponent $\eta=0.498(2)$ obtained with the algorithm for $N=1$ in 
two dimensions is in excellent agreement with the semi-exact result 
of $0.5$ inferred from \cite{Fis63}. Further, $\eta$ scales as $1/N$ for 
large $N$ as predicted in \cite{Wit78}.

We believe our algorithm can be useful in addressing some
interesting physics questions. For example, the universality of the finite 
temperature QCD chiral phase transition with staggered fermions 
remains controversial \cite{Aok98,Ber00}. It would be useful to give 
a definitive answer at least at strong couplings. Previous calculations 
could only be performed far from the chiral limit and on small lattices
\cite{Boy92}. Using our algorithm we can now settle this question 
completely. Another question of interest is to determine the 
critical density of baryons where chiral symmetry restoration takes 
place. At strong couplings and in the static approximation
(heavy baryon limit) this question can be answered with our algorithm.
Since baryons are known to be heavy at strong couplings this 
approximation may even be justifiable. Our algorithm can also be 
used to extract quantities like the decay width of the rho meson at 
strong couplings. This would be instructive since we would learn 
more about the difficulties involved in such calculations that are 
not merely algorithmic.

The present work also raises many interesting algorithmic questions.
Can one extend our algorithm to other situations? For example, can one
study more than one flavor of staggered fermions? This is interesting
since one can then explore more complex chiral symmetry breaking patterns.
Can Wilson fermions and Domain wall fermions be studied at strong 
couplings using similar techniques? Interesting 
and controversial phase structures have been predicted for the 
latter \cite{Gol00,Bro00}. What about a systematic way to go
towards the weak coupling limit? Perhaps allowing a fermionic 
plaquette term in the Boltzmann weight already has some desirable
effects. What about more general applications? For example it is
possible to rewrite the determinant of a matrix as a sum over 
monomer-dimer configurations. Does this mean we can calculate the
determinant of certain types of matrices more easily than before?

Finally, our algorithm should also be of interest to physicists studying
monomer-dimer systems. Such systems are interesting in their own right
and have a long history \cite{Hei72}. Many statistical mechanics problems 
including the Ising model can be reformulated in terms of monomer-dimer 
models on various kinds of lattices. Novel Monte Carlo algorithms continue 
to be developed to study them \cite{Kra02}. We believe that our approach 
can provide a useful alternative.

{\bf Acknowledgments}

We would like to thank Pierre van Baal, Wolfgang Bietenholz, Peter 
Hasenfratz, Ferenc Niedermayer, Costas Strouthos and Uwe-Jens Wiese 
for helpful comments. S.C. would also like thank U.-J. Wiese for 
hospitality during S.C.'s visit to Bern University where a major part 
of this work was done. This work was supported in part by funds 
provided by the U.S. Department of Energy grant DE-FG02-96ER40945. 
D.A. is supported at Leiden by a Marie Curie fellowship from the 
European Commission (contract HPMF-CT-2002-01716). The computations 
were performed on {\bf CHAMP}, (a 64-node Computer-cluster for Hadronic 
and Many-body Physics), funded by the Department of Energy and the 
Intel Corporation and located in the physics department at Duke
University.

\newpage

\appendix

\section{DERIVATION OF EXPRESSIONS FOR OBSERVABLES}
\label{obsder}

Here we present a brief derivation of the relations (\ref{optcorr}) and 
(\ref{tptcorr}). We begin by noting that the chiral condensate is given by
\be
\la\bpsi\psi(x)\ra=
\frac{1}{Z(m)}\,\sum_{\lb{}n,b\rb_x}\,W(\lb{}n,b\rb_x)\,m^{n}
\label{a1}
\ee
where the sum is over monomer-dimer configurations $\lb n,b\rb_x$ 
which satisfy the usual constraint relation (\ref{constraint}) on 
all sites except the site $x$ where the relation is modified to
\begin{equation}
n_x+\sum_{\mu}b_{\mu}(x)+b_{-\mu}(x)=N-1.
\end{equation} 
The partition function on the other hand is given by
\be
Z(m)\ =\ \sum_{\lb{}n,b\rb}\,W(\lb{}n,b\rb_x)\,m^{n}
\label{a2}
\ee
where the sum is over monomer-dimer configurations
$\lb n,b\rb$  which satisfy the usual constraint on all sites.
The weights $W(\lb n,b\rb_x)$ and $W(\lb n,b\rb)$ appearing in
(\ref{a1}) and (\ref{a2}) are obtained by multiplying the
block weights (\ref{actw}) and (\ref{pasw}) associated to all active and 
passive blocks. Since the mass is not taken into account in these
weights we have an extra factor $m^n$ where $n$ refers to the total 
number of monomers in the configuration.

Let us now construct an update to go from a given configuration 
$\lb n',b'\rb_x$ to a configuration $\lb n,b\rb$. This update 
is performed as follows: 
\begin{enumerate}
\item We declare the site $x$ to be passive.
\item We start at the site $x$ and choose the first bond to be one of 
the $2d$ bonds. The probability to choose the $\mu=\pm1$ bond is given
by $t^2/(2d-2+2t^2)$ and the probability to choose one of the 
$\mu \neq \pm 1$ bond by $1/(2d-2+2t^2)$.
\item Once we pick the bond we increase the dimer content of that link
by one and go to the adjoint active block.
\item We then use the probabilities discussed in section \ref{algo}
to construct a directed path update which ends on some active site $y$.
\end{enumerate}
Let the path generated using the above rules be referred to as 
$\alpha_{xy}$. At the end of the above update it is easy to prove that 
the new configuration belongs to the type $\lb n,b\rb$. Let 
$P(\lb n',b'\rb_x; \alpha_{xy})$ refer to the probability to produce 
the path $\alpha_{xy}$ starting from the configuration $\lb n',b'\rb_x$
using this update.

Now consider the unique reverse path of $\alpha_{xy}$ referred to as 
$\alpha^{-1}_{xy}$. This path is one of the ``partial'' directed paths 
that we produce using the directed path algorithm described in 
section \ref{algo}.
In particular this path is a path that starts from the active
site $y$ and visits the passive block at $x$. Let 
$P(\lb n,b\rb ; \alpha^{-1}_{xy})$ be the probability of 
generating this reverse path using the algorithm. Since the forward
and reverse paths are produced by probabilities that satisfy detailed 
balance at each stage except for possible factors at the ends, it
is easy to argue that
\begin{equation}
P(\lb n',b'\rb_x; \alpha_{xy}) \ W(\lb n',b'\rb_x) \ m^{n'} 
= \frac{V N m}{(2d-2+2t^2)} \ P(\lb n,b\rb ; \alpha^{-1}_{xy}) \ 
W(\lb n,b \rb) \ m^n
\label{db3}
\end{equation}
The factor $V$ arises because the site $y$ is picked with probability
$1/V$ but not the site $x$. The factor $N$ is due to the uncanceled factor
from (\ref{db2}). The factor $1/(2d-2+2t^2)$ compensates the
same factor from the left hand side. The mass factor arises because
of the mismatch in the number of monomers between the two 
configurations; in particular $n'=n+1$.

Now by construction we know
\begin{equation}
\sum_{\{\alpha_{xy}\}} P(\lb n',b'\rb_x; \alpha_{xy}) = 1
\end{equation}
where the sum is over all possible paths $\alpha_{xy}$. These paths
always start at the same $x$ but end at various sites $y$. Thus we
derive the relation
\begin{equation}
W(\lb n',b'\rb_x) m^{n'} 
= \sum_{\{\alpha_{xy}\}} \frac{V N m}{(2d-2+2t^2)} 
P(\lb n,b\rb ; \alpha^{-1}_{xy})\
W(\lb n,b \rb) m^n
\label{cwt}
\end{equation}
where the configurations $\lb n,b\rb$ on the right hand are determined
from the configuration $\lb n',b'\rb_x$ and the path $\alpha_{xy}$. It
is then possible to argue that
\begin{equation}
\sum_{\lb n',b'\rb_x} W(\lb n',b'\rb_x) m^{n'} 
= \sum_{\lb n,b\rb} 
\sum_{\{\alpha^{-1}_{xy}\}} \frac{V N m}{(2d-2+2t^2)} \ 
P(\lb n,b\rb; \alpha^{-1}_{yx})
W(\lb n,b \rb) m^n.
\end{equation}
Since the directed path update produces paths $\alpha^{-1}_{xy}$ starting
from the configuration $\lb n,b\rb$ with probability 
$P(\lb n,b\rb; \alpha^{-1}_{xy})$, we see that
\begin{equation}
\frac{1}{V}\la \bpsi\psi(x) \ra = \frac{m N}{2d-2+2t^2}
\Big\langle \sum_y I(x,y) \Big\rangle
\end{equation}
where $I(x,y)$ was defined in section \ref{obs} and the average
on the right hand side is taken over the ensemble of configurations
generated in the directed path algorithm. This proves relation
(\ref{optcorr}).

In order to show (\ref{tptcorr}) we start with the relation (\ref{cwt}) 
and assume that the mass factors are site dependent. By differentiating the
relation (\ref{cwt}) with respect to the mass factor at the site $y$ we 
can generate weights of configurations that contribute to the two monomer 
correlations. This then yields (\ref{tptcorr}). We leave the steps of this
derivation to the reader.

\section{EXACT RESULTS ON SMALL LATTICES}

In this appendix we give explicit expressions for the coefficients 
$c_{2n}$ defined in eq. (\ref{e2}), as a function of $x=t^2$  for 
a selection of small lattice sizes and small values of $N$ where exact 
calculations are possible.

\subsection{$N=1$ on a $2\times 2$ lattice}
\begin{eqnarray}
c_0 &=& 4(1 + x^2) \nonumber \\
c_2 &=& 4(1 + x) \nonumber \\
c_4 &=& 1 \nonumber
\end{eqnarray}

\subsection{$N=1$ on a $2\times 2\times 2$ lattice}
\begin{eqnarray}
c_0 &=& 16( 4 + 4 x^2 + x^4) \nonumber \\
c_2 &=& 8(16 + 16 x + 8 x^2 + 4 x^3) \nonumber \\
c_4 &=& 4 (20 + 16 x + 6 x^2) \nonumber \\ 
c_6 &=& 2 (8 + 4 x) \nonumber \\
c_8 &=& 1 \nonumber
\end{eqnarray}

\subsection{$N=1$ on a $4\times 4$ lattice}
\begin{eqnarray}
c_0 &=& 16(1+4x^2+7x^4+4x^6+x^8) \nonumber \\
c_2 &=& 64(2+4x+10x^2+13x^3+13x^4+10x^5+4x^6+2x^7) \nonumber \\
c_4 &=& 32(13+40x+81x^2+96x^3+81x^4+40x^5+13x^6) \nonumber \\
c_6 &=& 64(11+37x+63x^2+37x^4+11x^5) \nonumber \\
c_8 &=& 8(83+256x+354x^2+256x^3+83x^4) \nonumber \\
c_{10} &=& 32(11+28x+28x^2+11x^3) \nonumber \\
c_{12} &=& 8(13+24x+13x^2) \nonumber \\
c_{14} &=& 16(1+x) \nonumber \\
c_{16} &=& 1 \nonumber
\label{e3}
\end{eqnarray}

\subsection{$N=2$ on a $2\times 2\times 2$ lattice}
\begin{eqnarray}
c_0 &=& 1156+3136x^2+2116x^4+576x^6+81x^8 \nonumber \\
c_2 &=& 16(476+784x+1064x^2+870x^3+502x^4+264x^5+72x^6+27x^7) \nonumber \\
c_4 &=& 4(4428+8512x+9532x^2+6624x^3+3302x^4+1056x^5+243x^6) \nonumber \\
c_6 &=& 16(1132+2116x+1976x^2+1116x^3+386x^4+75x^5) \nonumber \\
c_8 &=& 2(4714+7744x+5768x^2+2304x^3+443x^4) \nonumber \\
c_{10} &=& 16(166+220x+116x^2+25x^3) \nonumber \\
c_{12} &=& 12(34+32x+9x^2)\nonumber \\
c_{14} &=& 16(2+x) \nonumber \\
c_{16} &=& 1 \nonumber
\label{e5}
\end{eqnarray}

\subsection{$N=1$ on a $4\times 2\times 2$ lattice}
\begin{eqnarray}
c_0 &=& 16(256+256x^2+112x^4+16x^6+x^8) \nonumber \\
c_2 &=& 128(128+128x+160x^2+104x^3+52x^4+20x^5+4x^6+x^7) \nonumber \\
c_4 &=& 32(832+1280x+1296x^2+768x^3+324x^4+80x^5+13x^6) \nonumber \\
c_6 &=& 64(352+592x+504x^2+252x^3+74x^4+11x^5) \nonumber \\
c_8 &=& 8(1328+2048x+1416x^2+512x^3+83x^4) \nonumber \\
c_{10} &=& 32(88+112x+56x^2+11x^3) \nonumber \\
c_{12} &=& 8(52+48x+13x^2) \nonumber \\
c_{14} &=& 16(2+x) \nonumber \\
c_{16} &=& 1 \nonumber
\label{e7}
\end{eqnarray}

\subsection{$N=1$ on a $2\times 2\times 4$ lattice}
\begin{eqnarray}
c_0 &=& 16(81+232x^2+216x^4+96x^6+16x^8) \nonumber \\
c_2 &=& 128(63+98x+142x^2+130x^3+84x^4+56x^5+16x^6+8x^7) \nonumber \\
c_4 &=& 32(563+1064x+1242x^2+944x^3+532x^4+192x^5+56x^6) \nonumber \\
c_6 &=& 64(284+529x+512x^2+312x^3+120x^4+28x^5) \nonumber \\
c_8 &=& 8(1179+1936x+1492x^2+640x^3+140x^4) \nonumber \\
c_{10} &=& 32(83+110x+60x^2+14x^3) \nonumber \\
c_{12} &=& 8(51+48x+14x^2) \nonumber \\
c_{14} &=& 16(2+x) \nonumber \\
c_{16} &=& 1 \nonumber
\label{e8}
\end{eqnarray}

\subsection{$N=2$ on a $4\times4$ lattice}

\begin{eqnarray}
c_0 &=& 65536(81+576x^2+2416x^4+5648x^6+7520x^8+5648x^{10} \nonumber \\
    && +2416x^{12}+576x^{14}+81x^{16}) \nonumber \\ 
c_2 &=& 2097152(54+144x+564x^2+1145x^3+2490x^4+3806x^5 \nonumber \\
    && +5470x^6+6303x^7+6303x^8+5470x^9+3806x^{10}+2490x^{11} \nonumber \\
    && +1145x^{12}+564x^{13}+144x^{14}+54x^{15})
\nonumber
\end{eqnarray}

\subsection{$N=3$ on a $2\times 2\times 2$ lattice}

\begin{eqnarray}
c_0 &=&{\textstyle \frac{1}{81}}
(4624+25600x^2+35396x^4+20224x^6+5800x^8+900x^{10}+81x^{12}) \nonumber \\
c_2 &=&{\textstyle \frac{2}{27}}
(10880+25600x+54080x^2+63556x^3+62912x^4+47560x^5 \nonumber \\
&&+28220x^6+15120x^7+5620x^8+2220x^9+450x^{10}+135x^{11}) 
\end{eqnarray}

\subsection{$N=2$ on a $4\times 2\times 2$ lattice}

\begin{eqnarray}
c_0 &=& 1336336+3625216x^2+5242816x^4+3817024x^6+1491668x^8 
\nonumber \\
&&
+318272x^{10}+37072x^{12}+2304x^{14}+81x^{16} 
\nonumber \\
c_2 &=& 64(275128+453152x+1053864x^2+1260996x^3+1522636x^4 
\nonumber \\
&&+1299408x^5+1009344x^6+630309x^7+334475x^8+154496x^9 
\nonumber \\
&&+56066x^{10}+19137x^{11}+4508x^{12}+1128x^{13}
\nonumber \\
&&+144x^{14}+27x^{15})
\end{eqnarray}

\subsection{$N=2$ on a $2\times 2\times 4$ lattice}

\begin{eqnarray}
c_0 &=& 198916+1682272x^2+4177396x^4+4825728x^6 \nonumber \\
    & & +3184704x^8+1343232x^{10}+381996x^{12}+69984x^{14}+6561x^{16} 
\nonumber \\
c_2 &=&32(140936+399424x+11511968x^2+1787069x^3+2560377x^4 \nonumber \\
&&+2707684x^5+2511784x^6+2002594x^7+1320438x^8+840288x^9 \nonumber \\
&&+403668x^{10}+212841x^{11}+70956x^{12}+31590x^{13}+5832x^{14} \nonumber \\
&&+2187x^{15})
\end{eqnarray}

\subsection{$N=1$ on a $2\times 2\times 2\times 2$ lattice}
\begin{eqnarray}
c_0 &=&  256(81+124 x^2+54 x^4+12 x^6+x^8) \nonumber \\
c_2 &=&  128(792+968x+936x^2+600x^3+240x^4 \nonumber \\
 && +144x^5+24 x^6+8x^7) \nonumber
\end{eqnarray}

\end{document}